\documentclass[apjl]{emulateapj}
\usepackage{comment}
\usepackage{ifthen}
\newcommand{\forloop}[5][1]
{
\setcounter{#2}{#3}
\ifthenelse{#4}
	{
	#5
	\addtocounter{#2}{#1}
	\forloop[#1]{#2}{\value{#2}}{#4}{#5}
	}
	{
	}
}

\newcommand{\lc}{light curve}
\newcommand{\lcs}{light curves}
\newcommand{\Lc}{Light curve}

\newcommand{\band}[1]{\ensuremath{#1}~band}

\newcommand{\kms}{\ensuremath{\rm km\,s^{-1}}}
\newcommand{\ms}{\ensuremath{\rm m\,s^{-1}}}

\newcommand{\gcmc}{\ensuremath{\rm g\,cm^{-3}}}
\newcommand{\ergscmsq}{\ensuremath{\rm erg\,s^{-1}\,cm^{-2}}}

\newcommand{\teff}{\ensuremath{T_{\rm eff}}}
\newcommand{\logg}{\ensuremath{\log{g}}}
\newcommand{\vsini}{\ensuremath{v \sin{i}}}
\newcommand{\feh}{\ensuremath{\rm [Fe/H]}}

\newcommand{\rsun}{\ensuremath{R_\sun}}
\newcommand{\msun}{\ensuremath{M_\sun}}
\newcommand{\lsun}{\ensuremath{L_\sun}}

\newcommand{\rstar}{\ensuremath{R_\star}}
\newcommand{\mstar}{\ensuremath{M_\star}}
\newcommand{\lstar}{\ensuremath{L_\star}}

\newcommand{\teffstar}{\ensuremath{T_{\rm eff\star}}}
\newcommand{\rhostar}{\ensuremath{\rho_\star}}
\newcommand{\loggstar}{\ensuremath{\log{g_{\star}}}}

\newcommand{\mearth}{\ensuremath{M_\earth}}

\newcommand{\rpl}{\ensuremath{R_{p}}}
\newcommand{\mpl}{\ensuremath{M_{p}}}

\newcommand{\rhopl}{\ensuremath{\rho_{p}}}

\newcommand{\arstar}{\ensuremath{a/\rstar}}
\newcommand{\zrstar}{\ensuremath{\zeta/\rstar}}

\newcommand{\rjup}{\ensuremath{R_{\rm J}}}
\newcommand{\mjup}{\ensuremath{M_{\rm J}}}

\newcommand{\rhojup}{\ensuremath{\rho_{\rm J}}}

\newcommand{\refsecl}[1]{\mbox{Section \ref{sec:#1}}}

\newcommand{\reftabl}[1]{Table~\ref{tab:#1}}

\newcommand{\hatcur}{HATS-17}
\newcommand{\hatcurb}{HATS-17b}
\newcommand{\hatcurRVgammaabs}{\hatcurRVgammaC}

\newcommand{\hatcurlumind}{\rhostar}
\newcommand{\hatcurjhkfilset}{ESO}

\newcommand{\hatcurSMEversion}{ii} 
\newcommand{\hatcurSMEteff}{\ifthenelse{\equal{\hatcurSMEversion}{i}}{\hatcurSMEiteff}{\hatcurSMEiiteff}}
\newcommand{\hatcurSMEzfeh}{\ifthenelse{\equal{\hatcurSMEversion}{i}}{\hatcurSMEizfeh}{\hatcurSMEiizfeh}}
\newcommand{\hatcurSMEzfehshort}{\ifthenelse{\equal{\hatcurSMEversion}{i}}{\hatcurSMEizfehshort}{\hatcurSMEiizfehshort}}
\newcommand{\hatcurSMElogg}{\ifthenelse{\equal{\hatcurSMEversion}{i}}{\hatcurSMEilogg}{\hatcurSMEiilogg}}
\newcommand{\hatcurSMEvsin}{\ifthenelse{\equal{\hatcurSMEversion}{i}}{\hatcurSMEivsin}{\hatcurSMEiivsin}}
\newcommand{\hatcurSMEvmac}{\ifthenelse{\equal{\hatcurSMEversion}{i}}{\hatcurSMEivmac}{\hatcurSMEiivmac}}
\newcommand{\hatcurSMEvmic}{\ifthenelse{\equal{\hatcurSMEversion}{i}}{\hatcurSMEivmic}{\hatcurSMEiivmic}}

\newcommand{\hatcurhtr}{HATS699-001} 
\newcommand{\hatcurCCra}{\ensuremath{12^{\mathrm h}48^{\mathrm m}45.72{\mathrm s}}}
\newcommand{\hatcurCCdec}{\ensuremath{-47{\arcdeg}36{\arcmin}49.3{\arcsec}}} 
\newcommand{\hatcurCCpmra}{\ensuremath{-32.40\pm0.90}} 
\newcommand{\hatcurCCpmdec}{\ensuremath{7.5\pm1.4}} 
\newcommand{\hatcurISOm}{\ensuremath{1.131\pm0.030}} 
\newcommand{\hatcurISOr}{\ensuremath{1.091_{-0.046}^{+0.070}}}
\newcommand{\hatcurLCPprec}{\ensuremath{16.2546107}} 
\newcommand{\hatcurLCPshort}{\ensuremath{16.2546}}
\newcommand{\hatcurPPr}{\ensuremath{0.777\pm0.056}}
\newcommand{\hatcurPPm}{\ensuremath{1.338\pm0.065}}
\newcommand{\hatcurXdist}{\ensuremath{340_{-17}^{+22}}}
\newcommand{\hatcurPPmlong}{\ensuremath{1.338\pm0.065}}
\newcommand{\hatcurPPrlong}{\ensuremath{0.777\pm0.056}} 
\newcommand{\hatcurRVeccen}{\ensuremath{0.029\pm0.022}}
\newcommand{\hatcurRVeccentwosiglim}{\ensuremath{<0.070}} 
\newcommand{\hatcurCCgsc}{GSC~8249-00170} 
\newcommand{\hatcurCCtwomass}{2MASS~12484555-4736492}
\newcommand{\hatcurSMEiteff}{\ensuremath{5840\pm91}}
\newcommand{\hatcurSMEizfeh}{\ensuremath{0.300\pm0.050}}
\newcommand{\hatcurSMEizfehshort}{\ensuremath{0.30}} 
\newcommand{\hatcurSMEilogg}{\ensuremath{4.36\pm0.15}}
\newcommand{\hatcurSMEivsin}{\ensuremath{3.84\pm0.48}}
\newcommand{\hatcurSMEivmac}{\ensuremath{0.0}}
\newcommand{\hatcurSMEivmic}{\ensuremath{0.0}}
\newcommand{\hatcurSMEiiteff}{\ensuremath{5846\pm78}}
\newcommand{\hatcurSMEiizfeh}{\ensuremath{0.300\pm0.030}}
\newcommand{\hatcurSMEiizfehshort}{\ensuremath{0.30}} 
\newcommand{\hatcurSMEiilogg}{\ensuremath{4\pm0}}
\newcommand{\hatcurSMEiivsin}{\ensuremath{3.73\pm0.39}}
\newcommand{\hatcurISOspec}{G} 
\newcommand{\hatcurRVgammaC}{\ensuremath{22943.2\pm4.0}}
\newcommand{\hatcurCCtassmB}{\ensuremath{13.105\pm0.090}}
\newcommand{\hatcurCCtassmv}{\ensuremath{12.39\pm0.10}}
\newcommand{\hatcurCCtassmg}{\ensuremath{12.665\pm0.050}}
\newcommand{\hatcurCCtassmr}{\ensuremath{12.162\pm0.060}}
\newcommand{\hatcurCCtassmi}{\ensuremath{12.08\pm0.19}}
\newcommand{\hatcurCCtwomassJmag}{\ensuremath{11.082\pm0.023}} 
\newcommand{\hatcurCCtwomassHmag}{\ensuremath{10.837\pm0.022}}
\newcommand{\hatcurCCtwomassKmag}{\ensuremath{10.698\pm0.021}}
\newcommand{\hatcurISOmlong}{\ensuremath{1.131\pm0.030}}
\newcommand{\hatcurISOrlong}{\ensuremath{1.091_{-0.046}^{+0.070}}}
\newcommand{\hatcurISOlogg}{\ensuremath{4.416\pm0.042}}
\newcommand{\hatcurLCrho}{\ensuremath{1.38\pm0.27}} 
\newcommand{\hatcurISOrho}{\ensuremath{1.22\pm0.17}}
\newcommand{\hatcurISOlum}{\ensuremath{1.24\pm0.17}} 
\newcommand{\hatcurISOmv}{\ensuremath{4.57\pm0.15}}  
\newcommand{\hatcurISOMK}{\ensuremath{3.08\pm0.12}} 
\newcommand{\hatcurISOage}{\ensuremath{2.1\pm1.3}} 
\newcommand{\hatcurXAv}{\ensuremath{0.17\pm0.11}}
\newcommand{\hatcurXdistred}{\ensuremath{339_{-16}^{+22}}}
\newcommand{\hatcurLCP}{\ensuremath{16.254611\pm0.000073}}
\newcommand{\hatcurLCT}{\ensuremath{2457139.1672\pm0.0014}}
\newcommand{\hatcurLCdur}{\ensuremath{0.2011\pm0.0038}}
\newcommand{\hatcurLCingdur}{\ensuremath{0.0166\pm0.0020}}
\newcommand{\hatcurPPar}{\ensuremath{25.8_{-1.5}^{+1.1}}}
\newcommand{\hatcurLCzeta}{\ensuremath{10.83\pm0.20}}
\newcommand{\hatcurLCrprstar}{\ensuremath{0.0726\pm0.0026}}
\newcommand{\hatcurLCbsq}{\ensuremath{0.187_{-0.083}^{+0.074}}}
\newcommand{\hatcurLCimp}{\ensuremath{0.432_{-0.110}^{+0.079}}}
\newcommand{\hatcurPPi}{\ensuremath{89.08\pm0.26}}
\newcommand{\hatcurLBii}{\ensuremath{0.2710}}
\newcommand{\hatcurLBiii}{\ensuremath{0.3370}}
\newcommand{\hatcurLBir}{\ensuremath{0.3640}} 
\newcommand{\hatcurLBiir}{\ensuremath{0.3272}}  
\newcommand{\hatcurRVK}{\ensuremath{99.1\pm4.4}} 
\newcommand{\hatcurRVrk}{\ensuremath{-0.037\pm0.087}}
\newcommand{\hatcurRVrh}{\ensuremath{-0.10_{-0.12}^{+0.16}}}
\newcommand{\hatcurRVjitterA}{\ensuremath{6\pm13}}
\newcommand{\hatcurRVjitterB}{\ensuremath{0.1\pm9.3}} 
\newcommand{\hatcurRVjitterC}{\ensuremath{0.9\pm4.1}}
\newcommand{\hatcurPPmrcorr}{\ensuremath{0.25}}  
\newcommand{\hatcurPPrho}{\ensuremath{3.50_{-0.51}^{+0.85}}}
\newcommand{\hatcurPPlogg}{\ensuremath{3.737_{-0.044}^{+0.060}}}
\newcommand{\hatcurPParel}{\ensuremath{0.1308\pm0.0012}}
\newcommand{\hatcurPPteff}{\ensuremath{814\pm25}} 
\newcommand{\hatcurPPtheta}{\ensuremath{0.398\pm0.031}} 
\newcommand{\hatcurPPfluxavgdim}{\ensuremath{7}}
\newcommand{\hatcurPPfluxavg}{\ensuremath{9.9\pm1.3}}

\newboolean{emulateapj}
\setboolean{emulateapj}{true}

\newboolean{rvtablelong}
\setboolean{rvtablelong}{true}

\newboolean{astroph}
\setboolean{astroph}{true}

\shortauthors{Brahm et al.}
\shorttitle{
\hatcur\lowercase{b}
}
\ifthenelse{\boolean{emulateapj}}{
    \newcommand{\titledag}{$\dagger$}
}{
    \newcommand{\titledag}{\dagger}
}

\begin{document}

\title{
\hatcur\lowercase{b}: A TRANSITING COMPACT WARM JUPITER IN A 16.3 DAYS CIRCULAR ORBIT
\altaffilmark{\titledag}
}

\author{
    R.~Brahm\altaffilmark{1,2},
    A.~Jord\'an\altaffilmark{1,2},
    G.~\'A.~Bakos\altaffilmark{3,$\star$,$\star\star$},
    K.~Penev\altaffilmark{3},
    N.~Espinoza\altaffilmark{1,2},
    M.~Rabus\altaffilmark{1,5},
    J.~D.~Hartman\altaffilmark{3},
    D.~Bayliss\altaffilmark{4},
    S.~Ciceri\altaffilmark{5},
    G.~Zhou\altaffilmark{6},
    L.~Mancini\altaffilmark{5},
    T.~G.~Tan\altaffilmark{7},
    M.~de~Val-Borro\altaffilmark{3},
    W.~Bhatti\altaffilmark{3},
    Z.~Csubry\altaffilmark{3},
    J.~Bento\altaffilmark{6},
    T.~Henning\altaffilmark{5},
    B.~Schmidt\altaffilmark{6},
    V.~Suc\altaffilmark{1},
    J.~L\'az\'ar\altaffilmark{8},
    I.~Papp\altaffilmark{8},
    P.~S\'ari\altaffilmark{8}
}
\altaffiltext{1}{Instituto de Astrof\'isica, Facultad de F\'isica,
  Pontificia Universidad Cat\'olica de Chile, Av. Vicu\~na Mackenna
  4860, 7820436 Macul, Santiago, Chile; rbrahm@astro.puc.cl}
\altaffiltext{2}{Millennium Institute of Astrophysics, Santiago, Chile}
\altaffiltext{3}{Department of Astrophysical Sciences, Princeton
                      University, NJ 08544, USA}
\altaffiltext{4}{Observatoire Astronomique de l'Universit\'e de Gen\`eve, 51 ch. des Maillettes, 1290 Versoix, Switzerland}
\altaffiltext{5}{Max Planck Institute for Astronomy, Heidelberg, Germany}
\altaffiltext{6}{Research School of Astronomy and Astrophysics, Australian National University, Canberra, ACT 2611, Australia}
\altaffiltext{7}{Perth Exoplanet Survey Telescope, Perth, Australia}
\altaffiltext{8}{Hungarian Astronomical Association, Budapest, Hungary}
\altaffiltext{$\star$}{Alfred P.~Sloan Research Fellow}
\altaffiltext{$\star\star$}{Packard Fellow}
\altaffiltext{$\dagger$}{
The HATSouth network is operated by a collaboration consisting of
Princeton University (PU), the Max Planck Institute f\"ur Astronomie
(MPIA), the Australian National University (ANU), and the Pontificia
Universidad Cat\'olica de Chile (PUC).  The station at Las Campanas
Observatory (LCO) of the Carnegie Institute is operated by PU in
conjunction with PUC, the station at the High Energy Spectroscopic
Survey (H.E.S.S.) site is operated in conjunction with MPIA, and the
station at Siding Spring Observatory (SSO) is operated jointly with
ANU.
This paper includes data gathered with the MPG~2.2\,m and ESO~3.6\,m
telescopes at the ESO Observatory in La Silla and with the 3.9\,m AAT in
Siding Spring Observatory.  This paper uses observations obtained with
facilities of the Las Cumbres Observatory Global Telescope.
}

\begin{abstract}
\setcounter{footnote}{10}
We report the discovery of \hatcurb{}, the first transiting {\em warm} Jupiter of the HATSouth network.
\hatcurb{} transits its bright (V=12.4) G-type (\mstar=\hatcurISOm\,\msun, \rstar=\hatcurISOr\,\rsun) metal-rich ([Fe/H]=$+0.3$ dex) host star in a circular orbit with a period of $P$=\hatcurLCPshort\ days. \hatcurb{} has a very compact radius of \hatcurPPr\,\rjup\ given its Jupiter-like mass of \hatcurPPm\,\mjup. Up to 50\% of the mass of \hatcurb{} may be composed of heavy elements in order to explain its high density with current models of planetary structure.
\hatcurb{} is the longest period transiting planet discovered to date by a ground-based photometric survey, and is one of the brightest transiting warm Jupiter systems known. The brightness of \hatcur{} will allow detailed follow-up observations to characterize the orbital geometry of the system and the atmosphere of the planet.
\setcounter{footnote}{0}
\end{abstract}


\section{Introduction}
\label{sec:introduction}

The detection of numerous extrasolar giant planets has brought forth several theoretical challenges
regarding their formation, structure and evolution.
One of these challenges arises from the fact that for over 20 years, radial velocity (RV) surveys have been discovering
large number of giant planets found to orbit their
host stars at short distances ($<1$ AU), where they are highly unlikely to be formed.
Hot Jupiters having semi-major axes of $\sim$0.03 AU, are the most extreme cases.
Short period giant planets are thought to be formed at several AUs, beyond the so-called snow line, where sufficient
solid material is available to build $\sim$10 \mearth\ cores that accrete their gaseous
envelopes from the protoplanetary disc before it is dispersed \citep[e.g.,][]{rafikov:2006}. The subsequent inward migration
can be produced by the exchange of angular momentum with the same protoplanetary disc \citep{goldreich:1980}
and/or by gravitational interactions with other stellar or planetary bodies which produce high eccentricity
migration mechanisms, in which eccentricities are excited and the semi-major axis decreases
due to tidal interactions with the star \citep{rasio:1996, wu:2011, fabrycky:2007, petrovich:2015}.
These two migration mechanisms predict different end products. While disc migration should
produce circular orbits in which the angular momentum vector of the orbit is aligned with the
spin of the star, high eccentricity migration mechanisms can produce planets with a broad distribution of eccentricities
and misalignment angles.

Transiting extrasolar planets (TEPs) are fundamental objects for constraining migration scenarios,
because the measurement of the Rossiter-McLaughlin effect \citep{rossiter:1924, mclaughlin:1924}
allows the determination of the sky-projected  angle between the orbital and stellar spins.
This angle has been determined for several transiting hot Jupiters showing that while most of the
systems have well aligned prograde orbits, an important fraction of them is found to present measurable
misalignments \citep{hebrard:2008, queloz:2010, winn:2010:RM}.
Hot Jupiters, however, are not optimal systems for discriminating between migration mechanisms.
The tidal or magnetic interactions with the host star which can arise due to their extremely close-in orbits
can be responsible for not only circularising the orbit but also potentially realigning the spin of the star with the orbit
of the planet and thereby affecting the final state of the system \citep{dawson:2014}.
Transiting giant planets with larger semi-major axes ($a>0.1$ AU), on the other hand, do not suffer
from strong interactions with their stars and can be used for measuring a more pristine final state of
the migration process.

While TEPs can be used to refine the geometrical configuration of the orbits, 
arguably their most important feature is that their radii can be derived from the transit
depth if the radii of the stellar hosts are known.
The estimation of the radius, coupled with the measurement of the planetary mass
from RV observations, allows the computation of the bulk density of the planet and
the possibility of inferring properties about its internal structure and composition.
Another theoretical challenge arose with the discovery of the first transiting
extrasolar planets. While theories of giant planet evolution predicted
 $\approx 1\rjup$ for planets with masses $\approx 1\mjup$, ages above 1Gyr and no cores \citep{burrows:2007},
observations of close-in transiting giant planets revealed a broad distribution
of planetary radii, with some of them reaching even twice the radius of Jupiter, like
HAT-P-32b \citep{hartman:2011:hat3233}. Others had radii more compact 
than expected from theoretical models without solid cores, like WASP-59b with $R_P$=0.78\rjup\
\citep{hebrard:2013:wasp59}.

The origin of these anomalies in the measured radii of giant exoplanets have been
extensively investigated, but there are no conclusive theories that are able to explain simultaneously
the variety of systems. A central solid core is commonly invoked
to explain the radii of compact giant planets, while the proximity of the planets
to their stellar hosts is probably responsible of generating the inflated planets via a variety of mechanisms including extra power deposited at some depth via, e.g., tidal or radiative heating mechanisms, enhanced atmospheric opacities, suppression of convective heat loss, among others \citep[for a review see][]{spiegel:2013}. The principal problem of favouring
one inflating mechanism over another comes from the degeneracies in the
modelled radius that arise from the unknown mass of the central core.
\cite{kovacs:2010} found that the inflation of the radius stops being efficient
for incident stellar fluxes weaker than $\langle F \rangle \approx 2 \times$10$^8$ erg s$^{-1}$ cm$^{-2}$ \citep[see also][]{demory:2011}.
Detections of giant planets with irradiation values below this limit are very valuable
because the interior structure of the planet can be estimated without assumptions about
extra energy sources. Furthermore, the distribution of core masses determined for weakly 
irradiated giant planets can then be extrapolated to highly irradiated
planets to constrain inflation mechanisms.

As stated above, giant TEPs with moderately long orbital periods (warm Jupiters) are unique test
objects for validating structure and migrations theories.
However, from the total of $\approx 1900$ confirmed or validated planets
discovered to date, only 23 transiting planets have $P>10$ days, $R_P>0.5$ \rjup\
and measured masses greater than 0.25 \mjup.
Moreover, most of these interesting systems were discovered with the space based
missions \textit{Kepler} and CoRoT around faint host stars ($V>14$)
hindering the measurement of precise RV variations, and limiting future detailed
follow-up observations.

On the other hand, the detection of transiting warm Jupiters from
the ground is a challenging task. Only two such planets with $P>10$ days are known,
originally discovered by RV programs and then later found to transit.
These are: HD17156 with $P=22.6$ days \citep{fisher:2007}
and HD80606 with $P=115$ days \citep{naef:2001}.
The small number of detections is due to the fact that the transit
probability decreases inversely with the semi-major axis.
Ground based transit surveys can deal in principle with this low
probability problem by monitoring many more stars than the RV programs do,
but the diurnal cycle strongly limits the recovery of $P>10$ days planets
for common one-site based surveys.
The use of longitudinal networks of telescopes is a way of counteracting the limitations imposed by the diurnal cycle.
Indeed, the transiting extrasolar planet with the longest period discovered previous to
the present study by a ground based transit survey was HAT-P-15b \citep{kovacs:2010} with $P=10.8$ days.
This system was detected by the two-site-based HATNet survey \citep{bakos:2004:hatnet}.

One of the main goals of the HATSouth survey  \citep{bakos:2013:hatsouth} is to expand the parameter space of well characterized transiting planets around moderately  bright stars. The first results in this regard have started to appear. HATS-6b
\citep{hartman:2014:hats6} is one of only four transiting giant planets discovered around
stars with masses $\mstar<0.6 \msun$, while HATS-7b \citep{bakos:2015:hats7} and HATS-8b \citep{bayliss:2015:hats8} are now two among the handful of well characterized transiting super Neptunes.
Having three locations with large longitude separation in the southern hemisphere,
the HATSouth survey is able to monitor, almost continuously, selected fields on the sky
for $\sim 4$ months per year, substantially increasing the probability of detecting transiting extrasolar planets
with periods longer than 10 days \citep{bakos:2013:hatsouth}. In this paper we present the discovery of \hatcurb{},
the first transiting warm Jupiter of the HATSouth survey. With an orbital period of $P\sim 16.3$ days, it
is the  longest period transiting extrasolar planet discovered by a ground-based
photometric survey.

The paper is structured as follows. In \S\ref{sec:obs} we describe the photometric and spectroscopic observations
that allowed the discovery and confirmation of \hatcurb{}. In \S\ref{sec:analysis}
we explain the tools that were used to estimate the physical parameters of \hatcurb{}
and its host star. Finally, in \S\ref{sec:discussion} we place the physical
properties of \hatcurb{} in the context of the transiting exoplanets previously detected
and outline possible follow-up observations for further characterizing this system.

\section{Observations}
\label{sec:obs}
\subsection{Photometry}
\label{sec:photometry}

\subsubsection{Photometric detection}
\label{sec:detection}

\begin{figure}[]
\plotone{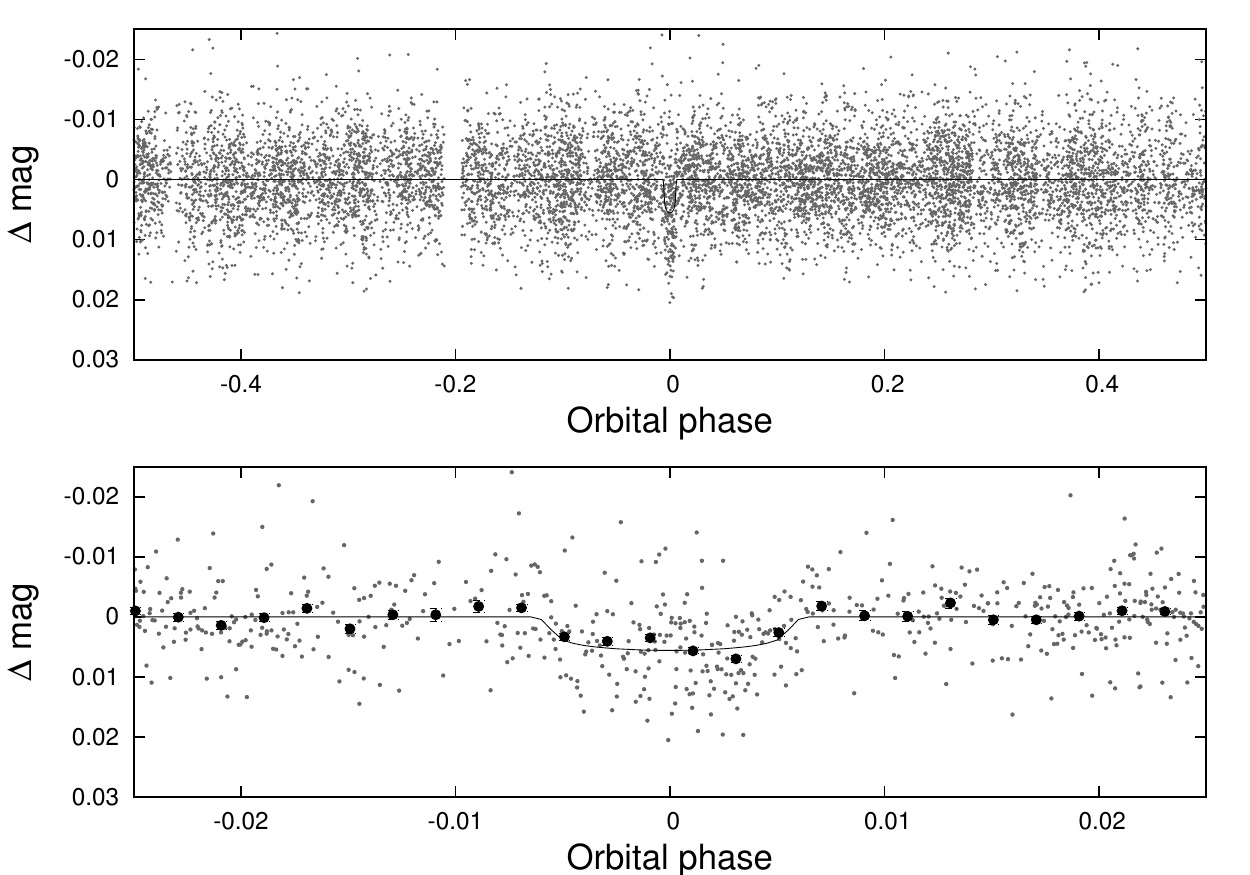}
\caption[]{
        Unbinned instrumental \band{r} \lc{} of \hatcur{} folded with
        the period $P = \hatcurLCPprec$\,days resulting from the global
        fit described in \refsecl{analysis}.  The solid line shows the
        best-fit transit model (see \refsecl{analysis}).  In the lower
        panel we zoom-in on the transit; the dark filled points here
        show the light curve binned in phase using a bin-size of 0.002.
        The signal is consistent with flat-bottom transit with a depth of 5 mmag.
        \\
\label{fig:hatsouth}}
\end{figure}

The star \hatcur{} (Table~\ref{tab:stellar}) was observed by HATSouth
instruments between UT 2011 April 26 and UT 2012 July 31 using the
HS-2, HS-4, and HS-6 units at LCO in Chile, the H.E.S.S.~site in
Namibia, and SSO in Australia, respectively. The number of
observations obtained with each instrument, effective cadence, and
photometric precision are listed in Table~\ref{tab:photobs}. The data
were reduced to trend-filtered light curves using the aperture
photometry procedure described by \citet{penev:2013:hats1} and making
use of External Parameter Decorrelation
\citep[EPD;][]{bakos:2010:hat11} and the Trend Filtering Algorithm
\citep[TFA;][]{kovacs:2005:TFA} to remove systematic variations.  We
searched for transits using the Box-fitting Least Squares
\citep[BLS;][]{kovacs:2002:BLS} algorithm, and detected a
$P=\hatcurLCPshort{}$\,day periodic transit signal in the light curve
of \hatcur{} (Figure~\ref{fig:hatsouth}; the data are available in
Table~\ref{tab:phfu}).

\ifthenelse{\boolean{emulateapj}}{
    \begin{deluxetable*}{llrrrr}
}{
    \begin{deluxetable}{llrrrr}
}
\tablewidth{0pc}
\tablecaption{
    Summary of Photometric Observations of \hatcur{}.
    \label{tab:photobs}
}
\tablehead{
    \colhead{Facility/Field \tablenotemark{a}} & 
    \colhead{Date Range} & 
    \colhead{Number of Points} & 
    \colhead{Median Cadence} & 
    \colhead{Filter} & 
        \colhead{Precision \tablenotemark{b}}\\
    \colhead{} & 
    \colhead{} & 
    \colhead{} &
    \colhead{(seconds)} &
    \colhead{} &
        \colhead{mmag}
}
\startdata
~~~~HS-2/G700 (LCO) & 2011 Apr--2012 Jul & 4579 & 292 & $r$ & 5.8 \\
~~~~HS-4/G700 (HESS) & 2011 Jul--2012 Jul & 3759 & 301 & $r$ & 6.4 \\
~~~~HS-6/G700 (SSO) & 2011 May--2012 Jul & 1499 & 300 & $r$ & 6.5 \\
~~~~PEST~0.3\,m (Perth, AU) & 2015 Apr 26 & 215 & 132 & $R_{C}$ & 2.4 \\
~~~~LCOGT~1\,m/sinistro (CTIO) & 2015 May 13 & 54 & 105 & $i$ & 1.3 \\
~~~~Swope~1\,m/e2v (LCO) & 2015 May 29 & 141 & 129 & $i$ & 3.8 \\
~~~~Swope~1\,m/e2v (LCO) & 2015 Jul 17 & 79 & 99 & $i$ & 1.6 \\
~~~~LCOGT~1\,m/sinistro (CTIO) & 2015 Jul 17 & 71 & 162 & $i$ & 0.8 \\
\enddata
\tablenotetext{a}{For the HATSouth observations we list the HS
  instrument used to perform the observations and the pointing on the
  sky. HS-2 is located at Las Campanas Observatory in Chile, HS-4 at
  the H.E.S.S. gamma-ray telescope site in Namibia, and HS-6 at Siding
  Spring Observatory in Australia. Field G700 is one of 838 discrete
  pointings used to tile the sky for the HATNet and HATSouth
  projects. This particular field is centered at R.A. 13.2\,hr and
  Dec.~$-45.0^{\circ}$.}
\tablenotetext{b}{The r.m.s.~scatter of the residuals from our best
  fit transit model for each light curve at the cadence indicated in
  the Table.}
\ifthenelse{\boolean{emulateapj}}{
    \end{deluxetable*}
}{
    \end{deluxetable}
}

\subsubsection{Photometric follow-up}
\label{sec:phfu}

Given the quality of the HATSouth detection, follow-up observations of the transit were
required in order to confirm that the signal is compatible with a transiting planet, and to
precisely determine the physical parameters of the system.

Due to the long period of the discovered transit signal and the 
long duration of the transits (5.2 hours), the photometric follow-up for this kind
of TEP candidate brings more difficulties than the ones presented in more typical
($P<5$\,days) candidates. For this reason a high priority photometric follow-up
campaign for \hatcur{} started only after the spectroscopic observations
described in Section~\ref{sec:hispec} showed an orbital variation in RV in phase
with the photometric ephemeris.

The first photometric follow-up light curve
of this system was obtained with the 0.3\,m Perth Exoplanet Survey
Telescope (PEST)\footnote{\url{http://www.cantab.net/users/tgtan/}} located
near Perth. The unbinned precision of 2.5 mmag allowed the measurement of
a full $\approx 5$ mmag flat-bottom transit.

Another two partial transits were then acquired with 
the LCOGT~1\,m telescope network, specifically with the telescope at Cerro Tololo
Inter-American Observatory (CTIO), and with the
Swope~1\,m coupled with the e2v camera at Las Campanas Observatory (LCO). The former registered only the
egress of the transit which was helpful in refining the ephemeris
of the system, while the latter obtained one ingress and part of the transit
but the weather conditions were suboptimal and did not allow for
a substantial improvement of the measured transit parameters.

Finally, two partial transits of the same event were measured
with high photometric precision ($\approx 1$ mmag) in one of the last chances to
observe it during the season. The observations were performed
with the same two telescopes that registered the previous partial transits
and they obtained a fraction of the bottom part of the transit and the egress.
These observations revealed a clear transit with a depth of $\approx 6$ mmag
and improved substantially the precision of the measured transit parameters.

\begin{figure*}[!ht]
\plotone{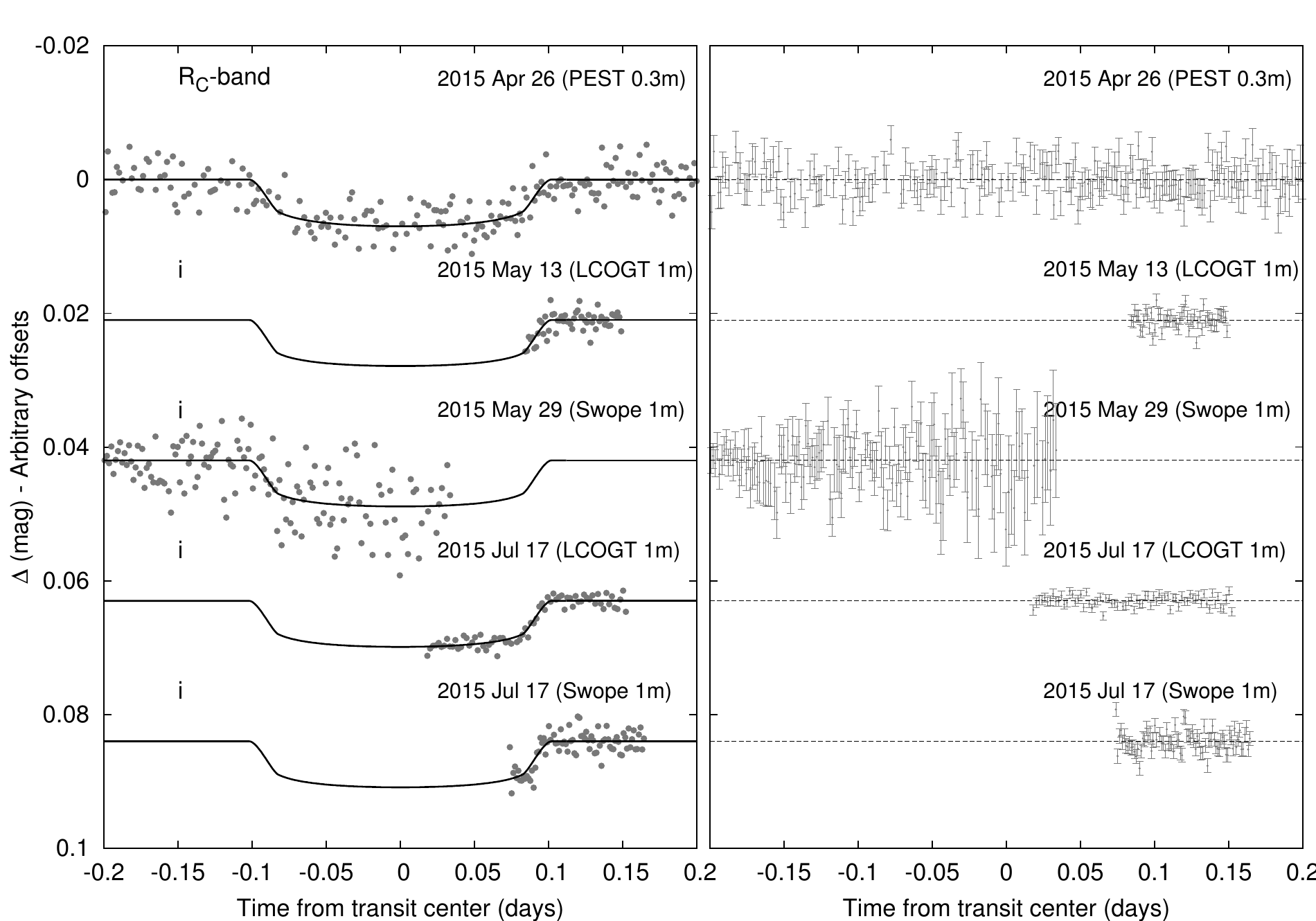}
\caption{
        Left: Unbinned follow-up transit \lcs{} of \hatcur{}.  The
        dates, filters and instruments used for each event are
        indicated.  Curves after the first are shifted for clarity.
        Our best fit is shown by the solid lines.  Right: Residuals
        from the fits in the same order as the curves at left.\\
\label{fig:lc}} \end{figure*}

All the photonetric observations are summarized in Table~\ref{tab:photobs}.
Table~\ref{tab:phfu} provides the light curve data, while the light curves
are compared to our best-fit model in Figure~\ref{fig:lc}.
All the facilities used for high precision photometric follow-up have been
previously used by HATSouth; the instrument specifications, observation
strategies and adopted reduction procedures can be found
in \cite{zhou:2014:mebs}, \cite{bakos:2015:hats7} and \cite{bayliss:2015:hats8}
for PEST, LCOGT~1\,m/sinistro (CTIO), and Swope~1\,m/e2v, respectively.
Given that there were no evident close companions to \hatcur{}, all
photometric follow-up observations were acquired with defocusing.

\ifthenelse{\boolean{emulateapj}}{
        \begin{deluxetable*}{lrrrrr} }{
        \begin{deluxetable}{lrrrrr} 
    }
        \tablewidth{0pc}
        \tablecaption{Differential photometry of
        \hatcur\label{tab:phfu}} \tablehead{ \colhead{BJD} &
        \colhead{Mag\tablenotemark{a}} &
        \colhead{\ensuremath{\sigma_{\rm Mag}}} &
        \colhead{Mag(orig)\tablenotemark{b}} & \colhead{Filter} &
        \colhead{Instrument} \\ \colhead{\hbox{~~~~(2\,400\,000$+$)~~~~}}
        & \colhead{} & \colhead{} & \colhead{} & \colhead{} &
        \colhead{} } \startdata
        $ 56123.25166 $ & $  -0.00660 $ & $   0.00285 $ & $ \cdots $ & $ r$ &         HS\\
        $ 56074.48974 $ & $   0.00167 $ & $   0.00334 $ & $ \cdots $ & $ r$ &         HS\\
        $ 56123.25650 $ & $   0.00042 $ & $   0.00287 $ & $ \cdots $ & $ r$ &         HS\\
        $ 56074.49324 $ & $   0.00716 $ & $   0.00339 $ & $ \cdots $ & $ r$ &         HS\\
        $ 56123.26027 $ & $  -0.00035 $ & $   0.00284 $ & $ \cdots $ & $ r$ &         HS\\
        $ 56074.49672 $ & $  -0.00865 $ & $   0.00337 $ & $ \cdots $ & $ r$ &         HS\\
        $ 56123.26373 $ & $  -0.00323 $ & $   0.00289 $ & $ \cdots $ & $ r$ &         HS\\
        $ 56074.50208 $ & $  -0.00251 $ & $   0.00346 $ & $ \cdots $ & $ r$ &         HS\\
        $ 56123.26876 $ & $  -0.00040 $ & $   0.00286 $ & $ \cdots $ & $ r$ &         HS\\
        $ 56074.50592 $ & $   0.00139 $ & $   0.00342 $ & $ \cdots $ & $ r$ &         HS\\
        [-1.5ex]

\enddata \tablenotetext{a}{
     The out-of-transit level has been subtracted. For the HATSouth
     light curve (rows with ``HS'' in the Instrument column), these
     magnitudes have been detrended using the EPD and TFA procedures
     prior to fitting a transit model to the light curve. Primarily as
     a result of this detrending, but also due to blending from
     neighbors, the apparent HATSouth transit depth is somewhat
     shallower than that of the true depth in the Sloan~$r$ filter
     (the apparent depth is 79\% that of the true depth). For the
     follow-up light curves (rows with an Instrument other than
     ``HS'') these magnitudes have been detrended with the EPD
     procedure, carried out simultaneously with the transit fit (the
     transit shape is preserved in this process).
}
\tablenotetext{b}{
        Raw magnitude values without application of the EPD
        procedure.  This is only reported for the follow-up light
        curves.
}
\tablecomments{
        This table is available in a machine-readable form in the
        online journal.  A portion is shown here for guidance
        regarding its form and content. The data are also available on
        the HATSouth website at \url{http://www.hatsouth.org}.
} \ifthenelse{\boolean{emulateapj}}{ \end{deluxetable*} }{ \end{deluxetable} }

\subsection{Spectroscopy}
\label{sec:hispec}

An extensive follow-up campaign is required for validating the
planetary nature of HATSouth transiting candidates. Transit-like
signals in the light curves can be produced by artifacts in the data or
different configurations of stellar eclipsing binaries and background stars.
Spectroscopic observations are used for characterizing the properties
of the star and to determine the mass and orbital parameters of the planets
from RV curves.

The first spectroscopic observation of \hatcur{} was carried out by the
WiFeS instrument on the ANU~2.3\,m telescope at SSO \citep{dopita:2007}.
A single low resolution (R=3000) spectrum was enough for
a first estimation of the stellar parameters of \hatcur{}.
Following the reductions and analysis procedures detailed
 in \cite{bayliss:2013:hats3}, the computed stellar atmospheric parameters
were $\teff=5315\pm300$ K, $\logg=4.4\pm0.3$ dex and $\feh=-0.5\pm0.5$ dex.
After \hatcurb{} was identified as a single-lined G-type dwarf, five additional R=7000 spectra
were obtained with the same instrument in order to measure RV variations. These five
RV points were consistent with no variation at the $\sim$2 \kms\ level, which shows
that the observed photometric signal is not produced by an unblended eclipsing stellar mass companion.
 
\begin{figure} [ht]
\plotone{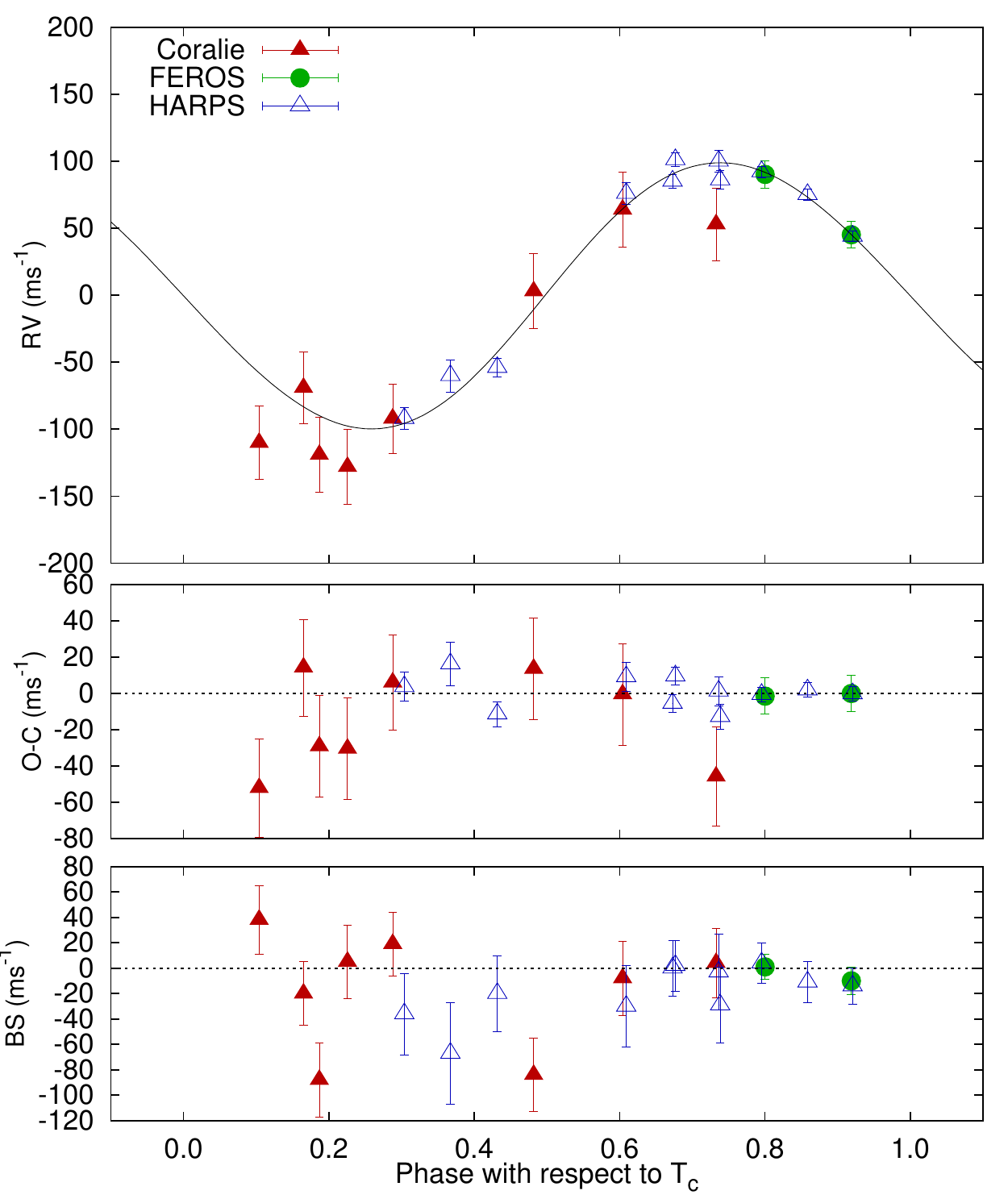}
\caption{
    {\em Top panel:} High-precision RV measurements from the
    MPG~2.2\,m/FEROS, the ESO~1.2\,m/CORALIE, and the ESO~3.6\,m/HARPS instruments,
    together with our best-fit orbit model.  Zero phase corresponds to
    the time of mid-transit.  The center-of-mass velocity has been
    subtracted.  {\em Second panel:} Velocity $O\!-\!C$ residuals from
    the best-fit model.  The error bars for each instrument include
    the jitter which is varied in the fit.  {\em Third panel:}
    Bisector spans (BS).  Note the different vertical scales of the
    panels.\\
\label{fig:rvbis}}
\end{figure}

 Once \hatcur{} passed the reconnaissance spectroscopy filter of our
follow-up structure, further spectroscopic characterisation of the \hatcur{} system was
performed with facilities capable of measuring RV
variations produced by the gravitational tug of a giant planet
mass companion. Several high resolution spectra were acquired with
three spectrographs installed in the ESO La Silla observatory. We
obtained 11 spectra using HARPS at the ESO~3.6\,m telescope, 8 spectra
using CORALIE \citep{queloz:2001} at the Euler~1.2\,m telescope and 2 spectra with
FEROS \citep{kaufer:1998} at the MPG~2.2\,m telescope. The data for these 3 instruments
were reduced and analysed with an automated pipeline described in \citet{jordan:2014:hats4}
that processes in an homogeneous
and robust manner data of echelle spectrographs. Besides the reduced
spectra, this pipeline delivers precise RV measurements, bisector span (BS)
values from the cross-correlation peak and an estimation of the
stellar atmospheric parameters. RV and BS values are presented in
Table~\ref{tab:rvs} with their corresponding uncertainties.
As shown in the top panel of Figure~\ref{fig:rvbis}, the RV measurements
phase cleanly with the photometric ephemeris with an amplitude compatible
with the one produced by a Jovian planet in an almost circular orbit.
The middle panel of Figure~\ref{fig:rvbis} shows the residuals
of the measured RV values and the best fit model, while the bottom panel
confirms that the BS values are not responsible for the measured RV variations.
The correlation coefficient between RV and BS values is $-0.2$ with a 95\% confidence interval
extending from $-0.596$ to $0.505$ obtained from a bootstrap simulation.
The mean atmospheric parameters obtained from the 3 spectrographs were:
$\teff=5705\pm118$ K, $\logg=4.14\pm0.25$ dex and $\feh=0.27\pm0.11$ dex,
where the errors in the parameters are computed from the dispersion of the 21
observations. These atmospheric parameters computed from high resolution
spectra show that \hatcur{} is significantly hotter and more metal rich than
we had previously estimated based on our initial WIFES spectrum.

\ifthenelse{\boolean{emulateapj}}{
    \begin{deluxetable*}{lrrrrrr}
}{
    \begin{deluxetable}{lrrrrrr}
}
\tablewidth{0pc}
\tablecaption{
    Relative radial velocities and bisector span measurements of
    \hatcur{}.
    \label{tab:rvs}
}
\tablehead{
    \colhead{BJD} & 
    \colhead{RV\tablenotemark{a}} & 
    \colhead{\ensuremath{\sigma_{\rm RV}}\tablenotemark{b}} & 
    \colhead{BS} & 
    \colhead{\ensuremath{\sigma_{\rm BS}}} & 
        \colhead{Phase} &
        \colhead{Instrument}\\
    \colhead{\hbox{(2\,456\,000$+$)}} & 
    \colhead{(\ms)} & 
    \colhead{(\ms)} &
    \colhead{(\ms)} &
    \colhead{} &
        \colhead{} &
        \colhead{}
}
\startdata
$ 1067.79423 $ & $    75.97 $ & $     8.00 $ & $  -30.0 $ & $   32.0 $ & $   0.609 $ & HARPS \\
$ 1068.83188 $ & $    84.97 $ & $     5.00 $ & $    0.0 $ & $   22.0 $ & $   0.673 $ & HARPS \\
$ 1068.89436 $ & $   100.97 $ & $     5.00 $ & $    2.0 $ & $   20.0 $ & $   0.677 $ & HARPS \\
$ 1069.85983 $ & $    99.97 $ & $     8.00 $ & $   -3.0 $ & $   30.0 $ & $   0.736 $ & HARPS \\
$ 1069.89514 $ & $    85.97 $ & $     7.00 $ & $  -29.0 $ & $   30.0 $ & $   0.738 $ & HARPS \\
$ 1070.82342 $ & $    91.97 $ & $     4.00 $ & $    4.0 $ & $   16.0 $ & $   0.795 $ & HARPS \\
$ 1071.84990 $ & $    74.97 $ & $     4.00 $ & $  -11.0 $ & $   16.0 $ & $   0.859 $ & HARPS \\
$ 1072.85895 $ & $    43.97 $ & $     4.00 $ & $  -14.0 $ & $   14.0 $ & $   0.921 $ & HARPS \\
$ 1075.83710 $ & $  -110.14 $ & $    17.00 $ & $   38.0 $ & $   27.0 $ & $   0.104 $ & Coralie \\
$ 1076.82996 $ & $   -69.14 $ & $    16.00 $ & $  -20.0 $ & $   25.0 $ & $   0.165 $ & Coralie \\
$ 1077.81340 $ & $  -128.14 $ & $    18.00 $ & $    5.0 $ & $   29.0 $ & $   0.226 $ & Coralie \\
$ 1078.82533 $ & $   -92.14 $ & $    15.00 $ & $   19.0 $ & $   25.0 $ & $   0.288 $ & Coralie \\
$ 1109.69531 $ & $  -119.14 $ & $    18.00 $ & $  -88.0 $ & $   29.0 $ & $   0.187 $ & Coralie \\
$ 1111.59481 $ & $   -92.03 $ & $     8.00 $ & $  -36.0 $ & $   32.0 $ & $   0.304 $ & HARPS \\
$ 1112.62535 $ & $   -60.03 $ & $    12.00 $ & $  -67.0 $ & $   40.0 $ & $   0.367 $ & HARPS \\
$ 1113.67401 $ & $   -54.03 $ & $     7.00 $ & $  -20.0 $ & $   30.0 $ & $   0.432 $ & HARPS \\
$ 1119.66549 $ & $    90.12 $ & $    10.00 $ & $    1.0 $ & $   10.0 $ & $   0.800 $ & FEROS \\
$ 1121.59323 $ & $    45.12 $ & $    10.00 $ & $  -10.0 $ & $   11.0 $ & $   0.919 $ & FEROS \\
$ 1179.50706 $ & $     2.86 $ & $    18.00 $ & $  -84.0 $ & $   29.0 $ & $   0.482 $ & Coralie \\
$ 1181.49387 $ & $    63.86 $ & $    18.00 $ & $   -8.0 $ & $   29.0 $ & $   0.604 $ & Coralie \\
$ 1183.58674 $ & $    52.86 $ & $    17.00 $ & $    4.0 $ & $   27.0 $ & $   0.733 $ & Coralie \\
    [-1.5ex]
\enddata
\tablenotetext{a}{
        The zero-point of these velocities is arbitrary. An overall
        offset $\gamma_{\rm rel}$ fitted separately to the data from
        three instruments has been subtracted.
}
\tablenotetext{b}{
        Internal errors excluding the component of
        astrophysical/instrumental jitter considered in
        \refsecl{analysis}.
}
\ifthenelse{\boolean{emulateapj}}{
    \end{deluxetable*}
}{
    \end{deluxetable}
}

\section{Analysis}
\label{sec:analysis}

\begin{figure}[]
\plotone{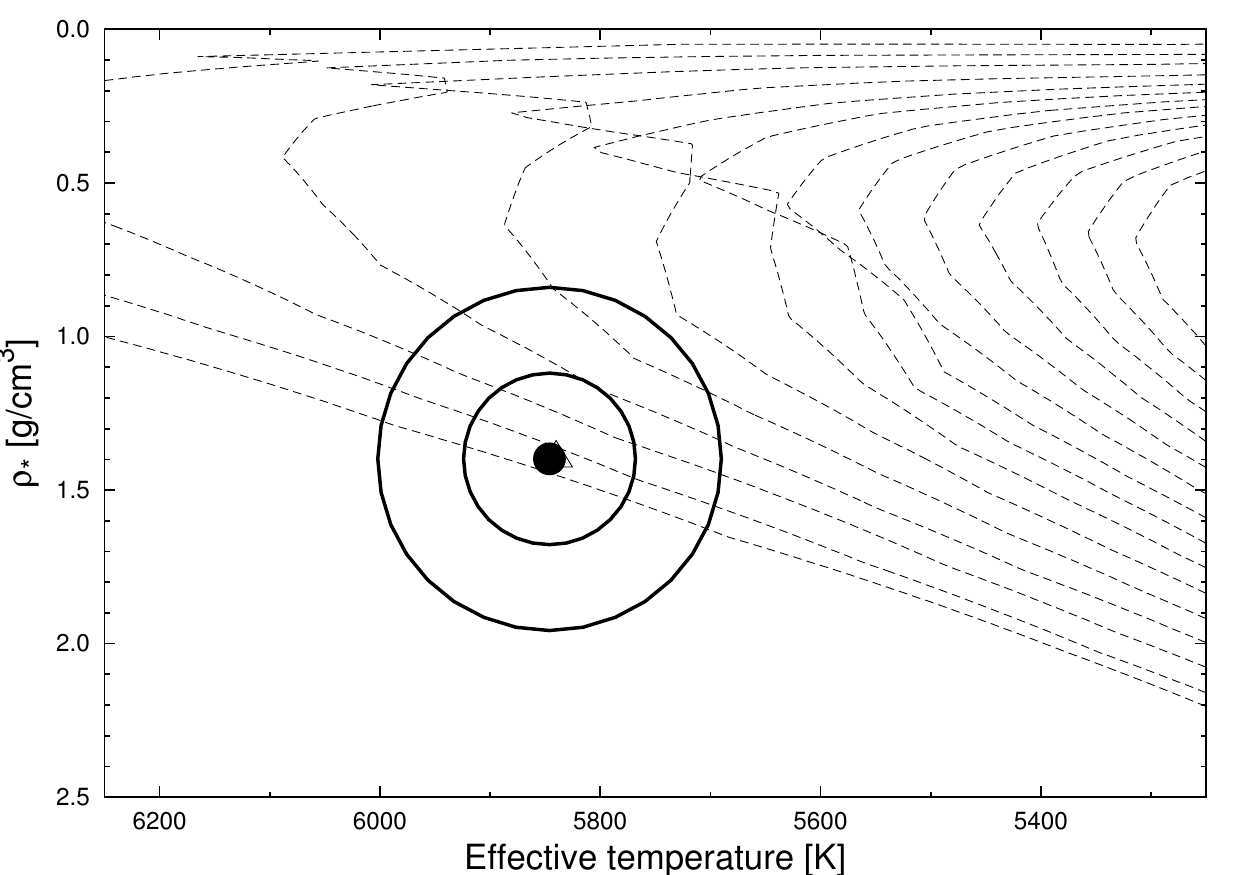}
\caption[]{
    Comparison between the measured values of \teffstar\ and
    \rhostar\ (from ZASPE applied to the FEROS spectra, and from our
    modeling of the light curves and RV data, respectively), and the
    Y$^{2}$ model isochrones from \citet{yi:2001}. The best-fit values
    (dark filled circle), and approximate 1$\sigma$ and 2$\sigma$
    confidence ellipsoids are shown. The values from our initial ZASPE
    iteration are shown with the open triangle. The Y$^{2}$ isochrones
    are shown for ages of 0.2\,Gyr, and 1.0 to 14.0\,Gyr in 1\,Gyr
    increments.
\label{fig:iso}}
\end{figure}

We analyzed the photometric and spectroscopic observations of
\hatcur{} to determine the parameters of the system using the standard
procedures developed for HATNet and HATSouth (see
\citealp{bakos:2010:hat11}, with modifications described by
\citealp{hartman:2012:hat39hat41}).

High-precision stellar atmospheric parameters were measured from the
FEROS spectra using the \texttt{ZASPE} code (Brahm et al. 2015, in prep). \texttt{ZASPE} estimates
the atmospheric stellar parameters and \vsini\ from high resolution
echelle spectra via a least squares method against a grid of synthetic
spectra in the most sensitive zones of the spectra to changes in the atmospheric
parameters. \texttt{ZASPE} obtains reliable errors in the parameters, as well
as the correlations between them by assuming that the principal
source of error is the systematic mismatch between the data and the
optimal synthetic spectra. We used a synthetic grid
provided by Brahm et al. (2015) and the spectral region considered
for the analysis was from 5000\AA\ to 6000\AA, which includes a large
number of atomic transitions and the pressure sensitive Mg Ib lines.  
We obtained the following high precision parameters with \texttt{ZASPE}:
\teffstar=5840$\pm$91 K, \loggstar=4.36$\pm$0.15 dex, \feh=0.30$\pm$0.05 dex
and $\vsini$=3.84$\pm$0.48 km/s.

The resulting \teffstar\ and \feh\ measurements were
combined with the stellar density \rhostar\ determined through our
joint light curve and RV curve analysis, to determine the stellar
mass, radius, age, luminosity, and other physical parameters, by
comparison with the Yonsei-Yale \citep[Y$^{2}$;][]{yi:2001} stellar
evolution models (see Figure~\ref{fig:iso}). This provided a revised
estimate of \loggstar\ which was fixed in a second iteration of
\texttt{ZASPE}. Our final adopted stellar parameters are listed in
Table~\ref{tab:stellar}; the final atmospheric parameters are compatible to the ones obtained in the first \texttt{ZASPE} iteration.
We find that the star \hatcur{} has a mass of
\hatcurISOm\,\msun, a radius of \hatcurISOr\,\rsun, and is at a
reddening-corrected distance of \hatcurXdist\,pc.

We also carried out a joint analysis of the high-precision FEROS,
CORALIE and HARPS RVs (fit using an eccentric Keplerian orbit) and the
HS, PEST~0.3\,m, LCOGT~1\,m, and Swope~1\,m light curves (fit using a
\citealp{mandel:2002} transit model with fixed quadratic limb
darkening coefficients taken from \citealp{claret:2004}) to measure
the stellar density, as well as the orbital and planetary
parameters. This analysis makes use of a differential evolution Markov
Chain Monte Carlo procedure \citep[DEMCMC;][]{terbraak:2006} to
estimate the posterior parameter distributions, which we use to
determine the median parameter values and their 1$\sigma$
uncertainties. The results are listed in
Table~\ref{tab:planetparam}. We find that the planet \hatcurb{} has a
mass of \hatcurPPmlong\,\mjup, and a radius of
\hatcurPPrlong\,\rjup. We also find that the observations are
consistent with a circular orbit: $e = \hatcurRVeccen$, with a 95\%
confidence upper-limit of $e\hatcurRVeccentwosiglim{}$. Note, however,
that due to the relatively long orbital period, and thus weak tidal
interaction between the planet and star, we do not fix the
eccentricity to zero in the fit, as we often do for shorter period
planets where there is a prior expectation of the orbit being
circular. The uncertainty in the eccentricity thus contributes to the uncertainties of other parameters listed in \reftabl{planetparam}.

In order to rule out the possibility that \hatcur{} is a blended stellar eclipsing binary system, we carried out a blend analysis of the photometric data following \citet{hartman:2012:hat39hat41}.
We find that all of the blend models considered provide a fit to the photometric data that has a higher $\chi^2$ than the model consisting of a single star with a transiting planet, and that the best-fitting blended eclipsing binary model can be rejected with $3\sigma$ confidence in favor of the single star with a planet model.  Moreover, the blend models which come closest to fitting the photometry would have easily been detected as a composite system based on the spectroscopic observations (RV and BS variations of several \kms). As is often the case, we find that while blends involving stellar eclipsing binaries may be ruled out by the photometry, we cannot exclude the possibility that \hatcur{} is a transiting planet system diluted by light from an unresolved stellar companion.  We find that including a physical wide binary companion with a mass $M > 0.5$\,\msun\ leads to a slightly higher $\chi^2$, but all companions, up to the mass of \hatcur{}, are permitted within $1\sigma$. If \hatcur{} has an unresolved stellar companion, the radius of \hatcurb{} could be as much as 1.6 times larger than what we infer here (for the extreme case of a star of equal mass to \hatcur{}).

Our analysis showed that \hatcur{} is a relatively young ($\sim$2 Gyr)
G-type star with physical parameters very similar to the Sun,
but with a substantial metal enrichment ([Fe/H]=$+0.3$).
On the other hand, \hatcurb{} is a weakly irradiated ($T_{\rm eq}\sim$800 K,
$\langle F \rangle\sim$10$^8$ erg s$^{-1}$cm$^{-2}$) Jovian planet and due
to its relatively long semi-major axis of $\sim$0.13 AU can be classified as a warm Jupiter.
One of the principal peculiarities of \hatcurb{} is that it has a very compact radius
for its Jupiter-like mass, yielding a very high density of $\rhopl$=3.5 gcm$^{-3}$ compared 
to Jupiter ($\rhojup$=1.33 gcm$^{-3}$) .

\ifthenelse{\boolean{emulateapj}}{
  \begin{deluxetable}{lcr}
}{
  \begin{deluxetable}{lcr}
}
\tablewidth{0pc}
\tabletypesize{\scriptsize}
\tablecaption{
    Stellar Parameters for \hatcur{} 
    \label{tab:stellar}
}
\tablehead{
    \multicolumn{1}{c}{~~~~~~~~Parameter~~~~~~~~} &
    \multicolumn{1}{c}{Value}                     &
    \multicolumn{1}{c}{Source}    
}
\startdata

\noalign{\vskip -3pt}

\sidehead{Identifying Information}
~~~~R.A.~(h:m:s)                      &  \hatcurCCra{} & 2MASS\\
~~~~Dec.~(d:m:s)                      &  \hatcurCCdec{} & 2MASS\\
~~~~R.A.p.m.~(mas/yr)                 &  \hatcurCCpmra{} & 2MASS\\
~~~~Dec.p.m.~(mas/yr)                 &  \hatcurCCpmdec{} & 2MASS\\
~~~~GSC ID                            &  \hatcurCCgsc{} & GSC\\
~~~~2MASS ID                          &  \hatcurCCtwomass{} & 2MASS\\

\sidehead{Spectroscopic properties}
~~~~$\teffstar$ (K)\dotfill         &  \hatcurSMEteff{} & ZASPE \tablenotemark{a}\\
~~~~Spectral type\dotfill           &  \hatcurISOspec & ZASPE\\
~~~~$\feh$\dotfill                  &  \hatcurSMEzfeh{} & ZASPE                 \\
~~~~$\vsini$ (\kms)\dotfill         &  \hatcurSMEvsin{} & ZASPE                 \\
~~~~$\gamma_{\rm RV}$ (\ms)\dotfill&  \hatcurRVgammaabs{} & FEROS                  \\

\sidehead{Photometric properties}
~~~~$B$ (mag)\dotfill               &  \hatcurCCtassmB{} & APASS                \\
~~~~$V$ (mag)\dotfill               &  \hatcurCCtassmv{} & APASS               \\
~~~~$g$ (mag)\dotfill               &  \hatcurCCtassmg{} & APASS                \\
~~~~$r$ (mag)\dotfill               &  \hatcurCCtassmr{} & APASS                \\
~~~~$i$ (mag)\dotfill               &  \hatcurCCtassmi{} & APASS                \\
~~~~$J$ (mag)\dotfill               &  \hatcurCCtwomassJmag{} & 2MASS           \\
~~~~$H$ (mag)\dotfill               &  \hatcurCCtwomassHmag{} & 2MASS           \\
~~~~$K_s$ (mag)\dotfill             &  \hatcurCCtwomassKmag{} & 2MASS           \\

\sidehead{Derived properties}
~~~~$\mstar$ ($\msun$)\dotfill      &  \hatcurISOmlong{} & Y$^{2}$+\hatcurlumind{}+ZASPE \tablenotemark{b}\\
~~~~$\rstar$ ($\rsun$)\dotfill      &  \hatcurISOrlong{} & Y$^{2}$+\hatcurlumind{}+ZASPE         \\
~~~~$\loggstar$ (cgs)\dotfill       &  \hatcurISOlogg{} & Y$^{2}$+\hatcurlumind{}+ZASPE         \\
~~~~$\rhostar$ (\gcmc)\dotfill       &  \hatcurLCrho{} & Light Curves         \\
~~~~$\rhostar$ (\gcmc)\dotfill  \tablenotemark{c}     &  \hatcurISOrho{} & $Y^{2}$+Light Curves         \\
~~~~$\lstar$ ($\lsun$)\dotfill      &  \hatcurISOlum{} & Y$^{2}$+\hatcurlumind{}+ZASPE         \\
~~~~$M_V$ (mag)\dotfill             &  \hatcurISOmv{} & Y$^{2}$+\hatcurlumind{}+ZASPE         \\
~~~~$M_K$ (mag,\hatcurjhkfilset{})&  \hatcurISOMK{} & Y$^{2}$+\hatcurlumind{}+ZASPE         \\
~~~~Age (Gyr)\dotfill               &  \hatcurISOage{} & Y$^{2}$+\hatcurlumind{}+ZASPE         \\
~~~~$A_{V}$ (mag) \tablenotemark{d}\dotfill           &  \hatcurXAv{} & Y$^{2}$+\hatcurlumind{}+ZASPE\\
~~~~Distance (pc)\dotfill           &  \hatcurXdistred{} & Y$^{2}$+\hatcurlumind{}+ZASPE\\
\enddata
\tablenotetext{a}{
    ZASPE = ``Zonal Atmospheric Stellar Parameter Estimator'' method
    for the analysis of high-resolution spectra (Brahm et al. 2015, in prep)
    applied to the FEROS spectra of \hatcur{}. These parameters rely
    primarily on ZASPE, but have a small dependence also on the
    iterative analysis incorporating the isochrone search and global
    modeling of the data, as described in the text.  }
\tablenotetext{b}{
    Isochrones+\hatcurlumind{}+ZASPE = Based on the Y$^{2}$ isochrones
    \citep{yi:2001},
    the stellar density used as a luminosity indicator, and the ZASPE
    results.
} 
\tablenotetext{c}{
   The stellar density as derived from the light curves, but also imposing a constraint that the combination of $T_{\rm eff}$, $\rhostar$ and [Fe/H] match to a stellar model from the $Y^2$ isochrones.
}
\tablenotetext{d}{ Total \band{V} extinction to the star determined
  by comparing the catalog broad-band photometry listed in the table
  to the expected magnitudes from the
  Isochrones+\hatcurlumind{}+ZASPE model for the star. We use the
  \citet{cardelli:1989} extinction law.  }
\ifthenelse{\boolean{emulateapj}}{
  \end{deluxetable}
}{
  \end{deluxetable}
}

\ifthenelse{\boolean{emulateapj}}{
  \begin{deluxetable}{lr}
}{
  \begin{deluxetable}{lr}
}
\tabletypesize{\scriptsize}
\tablecaption{Parameters for the transiting planet \hatcurb{}.
\label{tab:planetparam}}
\tablehead{
    \multicolumn{1}{c}{~~~~~~~~Parameter~~~~~~~~} &
    \multicolumn{1}{r}{Value \tablenotemark{a}}                     
}
\startdata
\noalign{\vskip -3pt}
\sidehead{\Lc{} parameters}
~~~$P$ (days)             \dotfill    & $\hatcurLCP{}$              \\
~~~$T_c$ (${\rm BJD}$)    
      \tablenotemark{b}   \dotfill    & $\hatcurLCT{}$              \\
~~~$T_{14}$ (days)
      \tablenotemark{b}   \dotfill    & $\hatcurLCdur{}$            \\
~~~$T_{12} = T_{34}$ (days)
      \tablenotemark{b}   \dotfill    & $\hatcurLCingdur{}$         \\
~~~$\arstar$              \dotfill    & $\hatcurPPar{}$             \\
~~~$\zrstar$ \tablenotemark{c}              \dotfill    & $\hatcurLCzeta{}$\phn       \\
~~~$\rpl/\rstar$          \dotfill    & $\hatcurLCrprstar{}$        \\
~~~$b^2$                  \dotfill    & $\hatcurLCbsq{}$            \\
~~~$b \equiv a \cos i/\rstar$
                          \dotfill    & $\hatcurLCimp{}$           \\
~~~$i$ (deg)              \dotfill    & $\hatcurPPi{}$\phn         \\

\sidehead{Limb-darkening coefficients \tablenotemark{d}}
~~~$c_1,i$ (linear term)  \dotfill    & $\hatcurLBii{}$            \\
~~~$c_2,i$ (quadratic term) \dotfill  & $\hatcurLBiii{}$           \\
~~~$c_1,r$               \dotfill    & $\hatcurLBir{}$             \\
~~~$c_2,r$               \dotfill    & $\hatcurLBiir{}$            \\

\sidehead{RV parameters}
~~~$K$ (\ms)              \dotfill    & $\hatcurRVK{}$\phn\phn      \\
~~~$e$ \tablenotemark{e}  \dotfill    & $\hatcurRVeccen{}$ \\
~~~$\sqrt{e}\cos \omega$   \dotfill    & $\hatcurRVrk{}$ \\
~~~$\sqrt{e}\sin \omega$   \dotfill    & $\hatcurRVrh{}$ \\
~~~Coralie RV jitter (\ms) \tablenotemark{f}        \dotfill    & \hatcurRVjitterA{}           \\
~~~FEROS RV jitter (\ms) \tablenotemark{f}        \dotfill    & \hatcurRVjitterB{}           \\
~~~HARPS RV jitter (\ms) \tablenotemark{f}        \dotfill    & \hatcurRVjitterC{}           \\

\sidehead{Planetary parameters}
~~~$\mpl$ ($\mjup$)       \dotfill    & $\hatcurPPmlong{}$          \\
~~~$\rpl$ ($\rjup$)       \dotfill    & $\hatcurPPrlong{}$          \\
~~~$C(\mpl,\rpl)$
    \tablenotemark{g}     \dotfill    & $\hatcurPPmrcorr{}$         \\
~~~$\rhopl$ (\gcmc)       \dotfill    & $\hatcurPPrho{}$            \\
~~~$\log g_p$ (cgs)       \dotfill    & $\hatcurPPlogg{}$           \\
~~~$a$ (AU)               \dotfill    & $\hatcurPParel{}$          \\
~~~$T_{\rm eq}$ (K) \tablenotemark{h}        \dotfill   & $\hatcurPPteff{}$           \\
~~~$\Theta$ \tablenotemark{i} \dotfill & $\hatcurPPtheta{}$         \\
~~~$\langle F \rangle$ ($10^{\hatcurPPfluxavgdim}$\ergscmsq) \tablenotemark{i}
                          \dotfill    & $\hatcurPPfluxavg{}$       \\ [-1.5ex]
\enddata
\tablenotetext{a}{
    For each parameter we give the median value and
    68.3\% (1$\sigma$) confidence intervals from the posterior
    distribution.
}
\tablenotetext{b}{
    Reported times are in Barycentric Julian Date calculated directly
    from UTC, {\em without} correction for leap seconds.
    \ensuremath{T_c}: Reference epoch of mid transit that
    minimizes the correlation with the orbital period.
    \ensuremath{T_{14}}: total transit duration, time
    between first to last contact;
    \ensuremath{T_{12}=T_{34}}: ingress/egress time, time between first
    and second, or third and fourth contact.
}
\tablenotetext{c}{
    Reciprocal of the half duration of the transit used as a jump
    parameter in our MCMC analysis in place of $\arstar$. It is
    related to $\arstar$ by the expression $\zrstar = \arstar
    (2\pi(1+e\sin \omega))/(P \sqrt{1 - b^{2}}\sqrt{1-e^{2}})$
    \citep{bakos:2010:hat11}.
}
\tablenotetext{d}{
    Values for a quadratic law, adopted from the tabulations by
    \cite{claret:2004} according to the spectroscopic (ZASPE) parameters
    listed in \reftabl{stellar}.
}
\tablenotetext{e}{
    While the eccentricity is allowed to vary in the fit, we find that
    the observations are consistent with a circular orbit.  The 95\%
    confidence upper-limit on the eccentricity is
    $e\hatcurRVeccentwosiglim$. We list $\sqrt{e}\cos\omega$ and
    $\sqrt{e}\sin\omega$ which are the jump parameters in the fit.
}
\tablenotetext{f}{
    Error term, either astrophysical or instrumental in origin, added
    in quadrature to the formal RV errors for the listed
    instrument. This term is varied in the fit assuming a prior inversely 
    proportional to the jitter.
}
\tablenotetext{g}{
    Correlation coefficient between the planetary mass \mpl\ and
    radius \rpl\ determined from the parameter posterior distribution
    via $C(\mpl,\rpl) = \langle(\mpl - \langle\mpl\rangle)(\rpl -
    \langle\rpl\rangle)\rangle/(\sigma_{\mpl}\sigma_{\rpl})\rangle$, 
	where $\langle \cdot \rangle$ is the
    expectation value operator, and $\sigma_x$ is the standard
    deviation of parameter $x$.
}
\tablenotetext{h}{
    Planet equilibrium temperature averaged over the orbit, calculated
    assuming a Bond albedo of zero, and that flux is reradiated from
    the full planet surface.
}
\tablenotetext{i}{
    The Safronov number is given by $\Theta = \frac{1}{2}(V_{\rm
    esc}/V_{\rm orb})^2 = (a/\rpl)(\mpl / \mstar )$
    \citep[see][]{hansen:2007}.
}
\tablenotetext{j}{
    Incoming flux per unit surface area, averaged over the orbit.
}
\ifthenelse{\boolean{emulateapj}}{
  \end{deluxetable}
}{
  \end{deluxetable}
}

\section{Discussion}
\label{sec:discussion}

\begin{figure*}[!ht]
\plotone{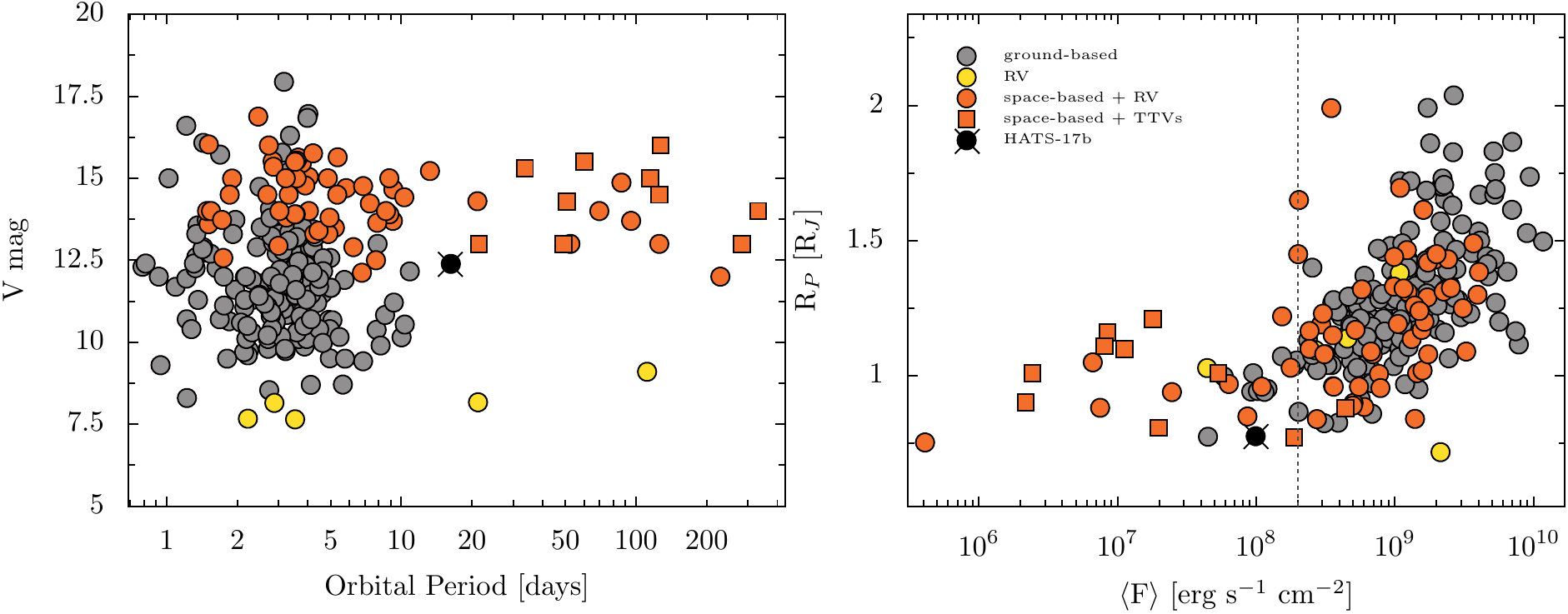}
\caption[]{
Left panel: V magnitude of the discovered transiting giant planets ($M_P$$>$0.25$M_J$, $R_P$$>$0.25$R_J$) as function of
their orbital periods. TEPs discovered by ground-based transit surveys correspond to the gray circles, while RV discovered
transiting planets a shown in yellow circles. TEPs discovered from space are shown in orange. Circles correspond to the planets
for which the masses determination were performed via RV measurements while squares are used to identify systems for which
the masses were estimates with TTVs. The black circle with a cross shows the position of \hatcurb{} which lies in a sparsely populated
region of the parameter space and stands out as the TEP with the longest orbital period discovered to date by a ground-based transit surveys.
Right panel: Radii of TEPs as function of the incoming flux per unit area in the surface of the planet. The symbols and colours represent
the same features as in the left panel. The vertical dashed line marks the insolation level below which extra heating mechanisms
do not produce inflated giant planets. \hatcurb{} lies in the zone of weakly irradiated planets.\\
\label{fig:periods}}
\end{figure*}

In this paper we have presented the discovery of \hatcurb{}, the first transiting warm Jupiter of the
HATSouth survey and the transiting extrasolar planet with the longest orbital period detected to date by
a ground-based photometric survey. The left panel of Figure~\ref{fig:periods} shows that \hatcurb{}, with
its period of 16.25 days, lies in a sparsely populated region of the parameter space of confirmed transiting
extrasolar giant planets ($M_p>0.25\,M_J$, $R_p>0.5\,R_J$) that have measured masses and densities.
There are only 19 confirmed giant planets with longer orbital periods, with most of them (17) discovered by the space-based
missions \textit{Kepler} and CoRoT, and orbiting stars that are generally too faint for performing detailed follow-up observations
to further characterize those systems. In fact, the masses of ten of the long period planets discovered from space
were determined by transit timing variations (TTVs) because it was easier than obtaining precise RV measurements of their faint host stars.
\hatcurb, on the other hand, has a bright ($V=12.4$) host which allowed a detailed determination of the orbital parameters
of the system via RV measurements and can be the target of future spectroscopic and photometric follow-up.

Due to its relatively large semi-major axis ($a\approx 0.13$ AU), \hatcurb{} is a low irradiated planet.
The flux received per unit area by the planet is $\langle F \rangle$=9.9$\times$10$^{7}$erg cm$^{-2}$s$^{-1}$,
which is low enough that we do not expect heating from the star to significantly impact the structure of the
planet \citep{kovacs:2010, demory:2011}. The right panel of Figure~\ref{fig:periods} shows that there are $\sim$30 other
well characterized giant planets with $\langle F \rangle$$<$2.0$\times$10$^{8}$erg cm$^{-2}$s$^{-1}$ that belong to the mentioned group,
with 11 of them discovered by ground-based transit surveys orbiting stars at shorter semi-major axes than \hatcurb{} but around less
luminous host stars than \hatcur{}.
In addition to the low insolation level of \hatcurb{}, the low eccentricity of its orbit ensures that tidal interactions
with the star are not able to generate internal heating on the planet. This particular state of \hatcurb{} is not
applicable for the whole group of low irradiated planets as many of them have measurable eccentricities
that could generate tidal heating during periastron passages.

\begin{figure*}[!ht]
\plotone{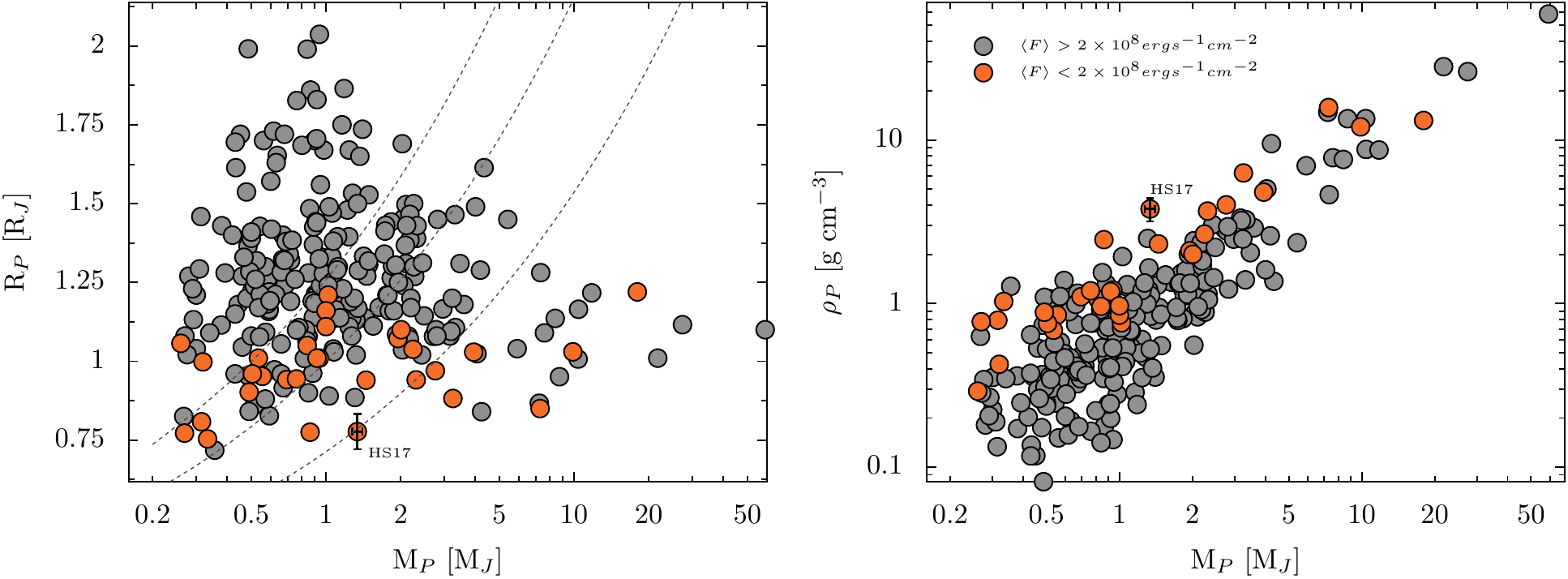}
\caption[]{
Left panel: mass-radius diagram of the detected giant TEPs. Black circles are planets with insulation levels greater than
$\langle F\rangle<2.0\times 10^{8}$ erg cm$^{-2}$s$^{-1}$, while orange circles correspond to planets receiving low irradiation.
\hatcurb{} present one of the smallest radii among transiting giant planets.
Right panel: density of giant planets as function of the planetary mass. \hatcurb{} lies at the upper envelope of this distribution.\\
\label{fig:dens}}
\end{figure*}

Transiting systems like \hatcurb{}, in which we can isolate the planetary physical properties from significant heating
mechanisms produced by the stellar host, are very important for constraining theoretical models of the structure of
giant planets. In Figure~\ref{fig:dens}, the physical properties of \hatcurb{} are contrasted with the ones of the
rest of the well characterized transiting giant exoplanets. Both panels illustrate that \hatcurb{} is a peculiar object
regarding its structure. \hatcurb{} possesses a radius of $R_P$=0.777\rjup\ which is extremely compact even for low
irradiated planets. The planet that most closely resembles \hatcurb{} is WASP-59b \citep{hebrard:2013:wasp59}
with $R_P$=0.78\rjup, $M_P$=0.86\mjup\ and $\langle F \rangle$=4.5$\times$10$^{7}$erg cm$^{-2}$s$^{-1}$.
The rest of the planets that share a similar radius have masses smaller than 0.4\mjup. The compact nature of
\hatcurb\  is further illustrated in the right panel of Figure~\ref{fig:dens},  where it stands out as the densest giant
planet with masses $M_P$$<$2\mjup.

\begin{figure}[]
\plotone{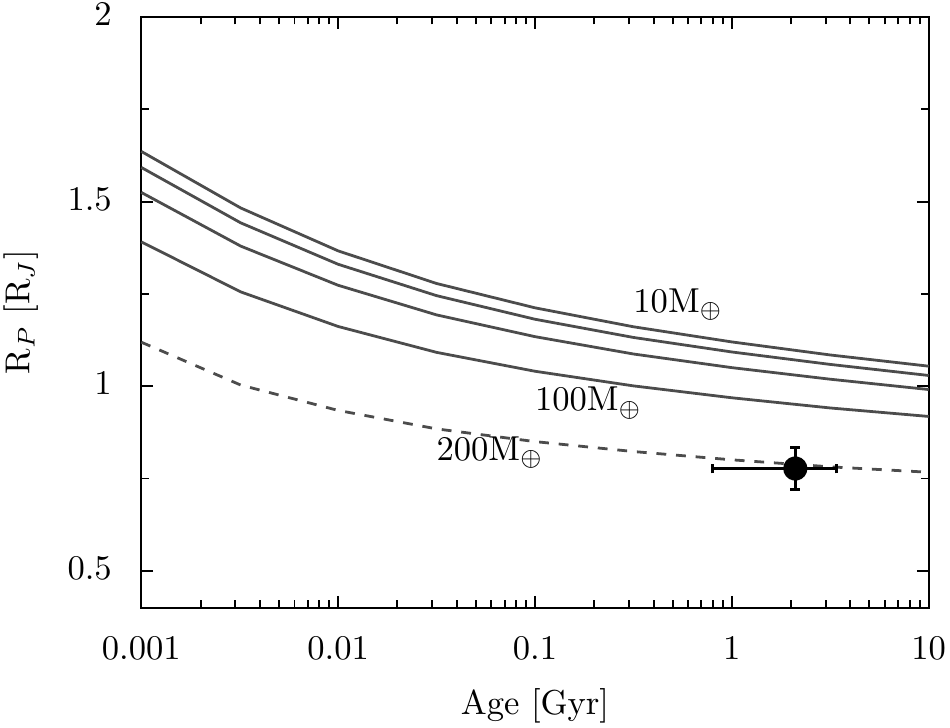}
\caption[]{
 Evolution models of the radius of a planet as a function of age for the planetary and stellar properties of the \hatcur{} system.
 The solid lines represent models with central core masses of 10, 25, 50 and 100\mearth\ from \cite{fortney:2007}. The
 dashed line corresponds to an extrapolation of the models for a core mass of 200\mearth. The filled circle corresponds to \hatcurb{}.
 A very high content of solid material is required to explain the compact radius of \hatcurb{}.\\
\label{fig:age-rad}}
\end{figure}

The small radius of \hatcurb{} is in concordance with the low irradiation levels of warm Jupiters.
However, its particular value is not straight-forward to explain with standard theoretical models of planetary structure.
Figure~\ref{fig:age-rad} shows that, for the stellar and planetary properties of the \hatcur{} system,
the \cite{fortney:2007}  models for giant planets predict a radius that is more than
3$\sigma$ larger than the observed one even for the maximum available core mass of 100\mearth.
By performing an extrapolation of the these models we have estimated
that a central core of $\sim$200\mearth\ is required to explain the compact nature of \hatcurb{}. Such a massive
core implies that $\sim$50\% of the planet mass is composed of heavy elements, which strongly contrasts with the $\sim$10\%
we can infer from Jupiter given a $\sim$15\mearth\ core \citep{militzer:2008} and is closer to the fraction of heavy elements predicted for the solar system ice giants.

\begin{figure*}[!ht]
\plotone{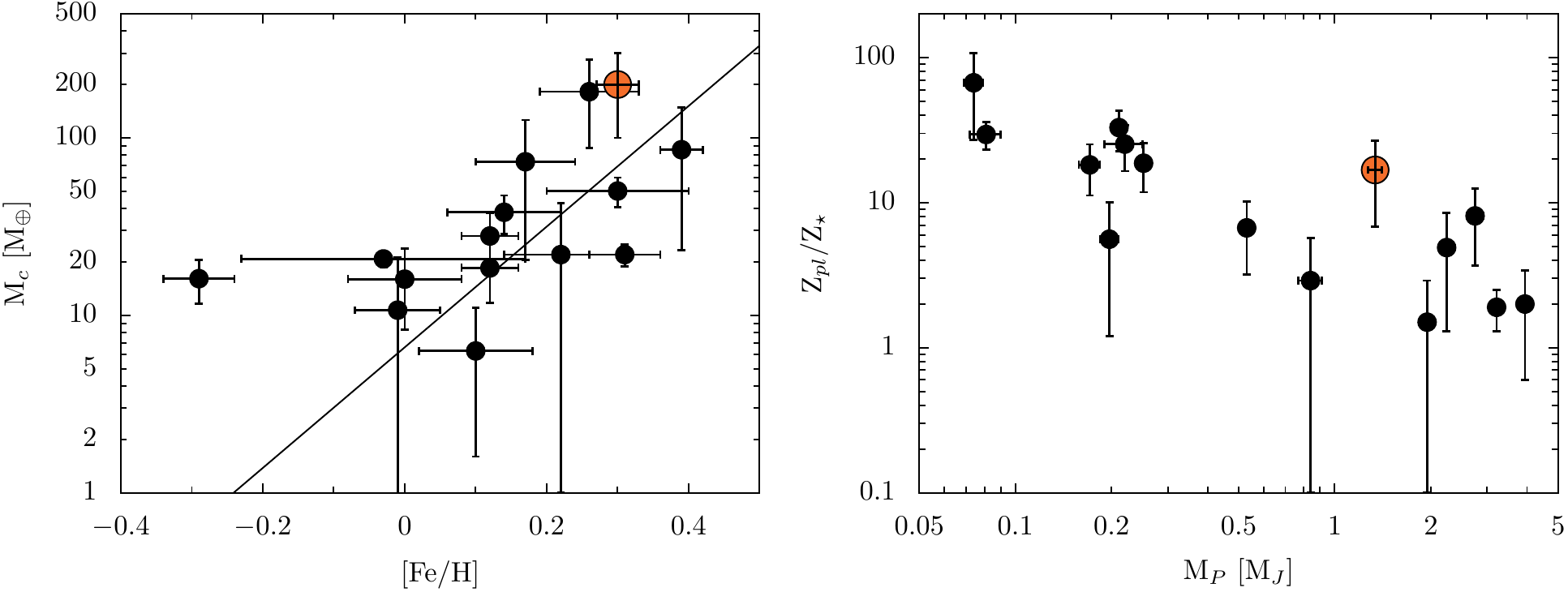}
\caption[]{
 Left: Correlation between the mass of the heavy elements in weakly irradiated planets and the metallicity of the host star. The gray circles
 correspond to the values found by \cite{miller:2011} and the orange circle represents \hatcur{}. The massive core of \hatcurb{} can be
 related to the high metallicity of its host.
 Right: Negative correlation proposed by  \cite{miller:2011} between the metal enrichment of the planets with respect to the metal enrichment of the
 star as a function of the mass of the planet. In this case \hatcurb{} tends to depart of the correlation and seems to lack of a massive H/He envelope
 for its given mass. \\
\label{fig:mcores}}
\end{figure*}

The massive core inferred for \hatcurb{} can be linked to the high metallicity of the parent star ([Fe/H]=$+0.3$ dex).
In the context of the core accretion scenario of giant planets formation, a more metal rich disk can be more
efficient in forming massive cores. Several works \citep{guillot:2006, burrows:2007, miller:2011} have claimed
to find a correlation between the stellar metallicity and the amount of heavy elements inferred for giant TEPs.
In particular, \citet{miller:2011} (hereafter M11) find that for low irradiated planets there is a minimum core mass of $\sim$10\mearth\
and that from this value the amount of heavy elements present in the planets' interior raises as a function of [Fe/H], with CoRoT-10b \citep{bonomo:2010}
being the most extreme case with a heavy element content of $M_c$=182$\pm$94\mearth\ and [Fe/H]=$+0.26$ dex.
The left panel of Figure~\ref{fig:mcores} shows this claimed correlation for the 14 systems analyzed by M11 and adding \hatcurb{}. \hatcurb{} seems to agree quite well with the correlation proposed by M11. Even though
the predicted heavy element content for a metallicity of [Fe/H]=$+0.3$ dex should be closer to $\sim$100\mearth, the dispersion
of the correlation is greater than the individual errors. Clearly, detections of more warm giant TEPs are required. 

M11 also proposed a negative correlation between the metal enrichment of the {\em planet} relative to the star and the mass of the planet.
This correlation is observed in the giant planets of our solar system where Uranus and Neptune are more enriched in heavy
elements than Saturn and in turn Saturn is more metal enriched than Jupiter. The right panel of Figure~\ref{fig:mcores} shows
that \hatcurb{} seems to subtly depart from this correlation having enrichments similar to Saturn mass planets rather than the ones
of Jupiter mass planets.

In summary, the massive core of \hatcurb{} can be expected given the high metallicity of the parent star,
but it seems to lack a more extended H/He envelope.
The mechanism that allows the formation of such massive embryos is unclear. If \hatcurb{} was formed
by core accretion at $a=5$ AU and we assume 
that the total heavy element composition of \hatcur{} scales with the iron abundance, we can infer an
embryo of $M_{c}$=30\mearth , which corresponds to just 15\% of the estimated mass of the core of \hatcurb.
More massive cores can be formed at larger distances but even if the primordial material is available, the planetesimal
accretion rate must exceed the gas accretion rate which should be difficult to accomplish for cores with $M_c>$20\mearth.
An alternative explanation for the extremely massive core of \hatcurb{} can be related to collisions with other objects in the system
posterior to the dispersal of the protoplanetary disk.
\cite{liu:2015} proposed, based on numerical simulations, that giant impacts of super-Earth-like planets or
mergers with other gas giants generally leads to a total coalescence of impinging gas giants and that sometimes
the collisions can disintegrate the envelope of gas giants which may also explain the seeming lack of a massive H/He envelope for \hatcurb{}.
This hypothesis is further supported by the study of \cite{petrovich:2014} which determined that at small semi-major axes ($a<0.5$ AU),
gravitational interactions between planets in unstable systems mostly lead to collisions rather than excitation of highly eccentric and inclined planetary orbits. 

A more detailed modelling of the structure of \hatcurb{}, in which the solid material is distributed through the entire envelope of the planet
and not only in a central core, can also lower the amount of heavy elements required to explain its small radius.
For example, in the case of the massive planet CoRoT-20b \citep[$M_P=4.24 \mjup$]{deleuil:2012}, the inclusion of the \cite{baraffe:2008}
calculations can decrease by a factor of three the 800\mearth\ in heavy elements that were initially estimated for this planet.\\

The current orbital distance of \hatcurb{} from its host star is compatible with migration via angular momentum exchange with
the protoplanetary disk. Migrations through gravitational interactions with other planetary and/or stellar companions should
excite the eccentricity of the system and then tidal interactions with the star during periastron passages would be responsible
of decreasing the semi-major axis and circularising the orbit. The eccentricity of \hatcurb{} is consistent with $e=0$, while being
too far away from it parent star to have suffered from significant tidal interactions. On the other hand, disc migration is expected to suppress
any initial eccentricity of giant planets \citep{dunhill:2015}.
While disk migration stands up as the most probable origin for the current semi-major axis of \hatcurb{}, high eccentricity
migration mechanisms cannot be totally discarded. Kozai-Lindov oscillations \citep{kozai:1962} produced by interactions
with a distant stellar companion \citep{takeda:2005} or with a closer planetary companion \citep{naoz:2011} can be taking
place but we may be just observing a stage of low eccentricity in the cycle.
Long term RV monitoring of \hatcurb{} can unveil the presence of another object in the system and measurements of the
Rossiter-Mclaughlin effect can detect inclinations in the orbit of \hatcurb{} produced by the interaction with the companion.
However, \cite{dong:2014} predicts that non-eccentric warm Jupiters probably present well aligned orbits with the spin of the star. \\

As evident from the previous paragraphs, \hatcurb{} belongs to a group of exoplanets that are useful for constraining
theories of structure and evolution of giant planets, but which has a  low number of well characterized systems discovered to date.
The detection of these transiting warm Jupiters around bright stars is fraught with several difficulties due to the low transit probability of
long period planets, low occurrence rate of giant planets with respect to terrestrial planets, and the limited duty cycle that one site
ground-based transit surveys are affected by. Moreover the confirmation of the planetary nature of transiting warm Jupiter candidates
requires extensive spectroscopic and photometric follow-up campaigns in which observations must be spread over many more epochs
compared to the follow-up observations required to confirm short period planets. Taking advantage of its three observing sites in the
southern hemisphere, separated by almost 120 deg in longitude each, the HATSouth survey can better tackle these difficulties, and \hatcurb{} is a testament to its capabilities.

\acknowledgements 

\paragraph{Acknowledgements}
Development of the HATSouth project was funded by NSF MRI grant
NSF/AST-0723074, operations have been supported by NASA grants
NNX09AB29G and NNX12AH91H, and follow-up observations received partial
support from grant NSF/AST-1108686.
R.B.\ and N.E.\ are supported by CONICYT-PCHA/Doctorado
Nacional. 
A.J.\ acknowledges support from FONDECYT project 1130857, BASAL CATA
PFB-06, and from the Ministry of Economy, Development, and Tourism's Millennium Science Initiative through grant IC120009, awarded to The Millennium Institute of Astrophysics, MAS.  
R.B.\ and N.E.\ acknowledge additional support rom the Ministry of Economy, Development, and Tourism's Millennium Science Initiative through grant IC120009, awarded to The Millennium Institute of Astrophysics, MAS.
V.S.\
acknowledges support form BASAL CATA PFB-06.  
This work is based on observations made with ESO Telescopes at the La
Silla Observatory.
This paper also uses observations obtained with facilities of the Las
Cumbres Observatory Global Telescope.
Work at the Australian National University is supported by ARC Laureate
Fellowship Grant FL0992131.
We acknowledge the use of the AAVSO Photometric All-Sky Survey (APASS),
funded by the Robert Martin Ayers Sciences Fund, and the SIMBAD
database, operated at CDS, Strasbourg, France.
Operations at the MPG~2.2\,m Telescope are jointly performed by the
Max Planck Gesellschaft and the European Southern Observatory.  The
imaging system GROND has been built by the high-energy group of MPE in
collaboration with the LSW Tautenburg and ESO\@.  
G.~B.~wishes to thank the warm hospitality of Ad\'ele and Joachim Cranz
at the farm Isabis, supporting the operations and service missions of
HATSouth. 

\clearpage
\bibliographystyle{apj}
\bibliography{hatsbib}

\end{document}